# Customer Selection Model with Grouping and Hierarchical Ranking Analysis

Bowen Cai

Marketing Analytics Assistant,

The Wharton School, University of Pennsylvania

8/15/2017

**Abstract**

The purpose of this study was to build a customer selection model based on 20 dimensions, including customer codes, total contribution, assets, deposit, profit, profit rate, trading volume, trading amount, turnover rate, order amount, withdraw amount, withdraw rate, process fee, process fee submitted, process fee retained, net process fee retained, interest revenue, interest return, exchange house return I, exchange house return II, to group and rank customers. The traditional way to group customers in securities/futures companies is simply based on their assets. However, grouping customers with respect to only one dimension cannot give us a full picture about customers' attributions. It is hard to group customers' with similar attributions or values into one group if we just consider assets as the only grouping criterion. Nowadays, securities/futures companies usually group customers based on managers' experience with lack of quantitative analysis, which is not effective. Therefore, we use kmeans unsupervised learning methods to group customers with respect to significant dimensions so as to cluster customers with similar attributions together.

Grouping is our first step. It is the horizontal analysis in customer study. The next step is customer ranking. It is the longitudinal analysis. It ranks customers by assigning each customer with a certain score given by our weighted customer value calculation formula. Therefore, by grouping and ranking customers, we can differentiate our customers and rank them based on values instead of blindly reaching everyone.

## Introduction

In recent years, China Securities Regulatory Commission (CSRC) requires securities companies to properly group and classify customers. According to the new regulation in 2010, securities companies should follow up customers more frequently, which is no less than 10% of the total number of customers in the end of last year. Many big securities companies have tried to use statistical methods to group and rank their customers, but the results were not satisfactory. Many common statistical methods that can be applied in this problem are linear regression, logistic regression, clustering, decision tree, random forest, gradient boosting method and support vector machine. In this paper, we apply clustering method in horizontal analysis and gradient boosting method in longitudinal analysis.

Clustering is the partitioning or grouping process in a given set of patterns to produce disjoint clusters (Alsabti, Ranka & Singh, 2002). This algorithm works iteratively to assign each data point to one of k groups based on feature similarity. We can get the centroids of the k clusters and labels for the training data, since each data point is assigned to a single cluster. However, choosing the number of k is a tricky part in this algorithm. Normally speaking, k can be specified by the user with reference to the background information. Nevertheless, we can use elbow method to identify the number of clusters by observing the change of the direction of the cost function (Gorakala S.K. & Usuelli M. (2015) *Building a Recommendation System with R.* Birmingham, UK: Packt Publishing Ltd.). In our case, we use elbow method to identify the number of clusters and compare the answer to our experience. The two answers are the same.

We use stochastic gradient boosting in customer ranking analysis. Stochastic gradient boosting relies on an empirical approximation of the true gradient with associated with a particular kind of regression mean function: linear regression, logistic regression, robust regression, Poisson regression, quantile regression, and others (Berk, R.A. (2016) *Statistical Learning from a Regression Perspective.* Philadelphia, PA: Springer). With stochastic gradient boosting, each tree is constructed much as a conventional regression tree (Berk, R.A. (2016) *Statistical Learning from a Regression Perspective.* Philadelphia, PA: Springer). The difference exists in the definition for the fitting target. Usually, we consider random forest as "big trees" because it requires plenty of trees in the process, but we think gradient boosting methods as "small trees" because we only need a number of trees that is enough to observe the changes in the loss function with OOB (Out of Bagging) data to get the stopping point of the iteration.

In addition, we can use interaction plot to decide the relationship between two particular dimensions, for example, assets vs. contribution. It is helpful for securities companies to gain insightful findings into customer values.

This paper will be divided into two parts. The first part will introduce horizontal analysis in customer grouping. The second part will introduce longitudinal analysis in customer ranking. The quantitative analysis is developed in the R environment, and the dataset is from an anonymous big futures company in China.

## Customer Grouping

Before applying clustering methods, we need to reduce the dimensionality. We have 20 dimensions, including customer codes, total contribution, assets, deposit, profit, profit rate, trading volume, trading amount, turnover rate, order amount, withdraw amount, withdraw rate, process fee, process fee submitted, process fee retained, net process fee retained, interest revenue, interest return, exchange house return I, exchange house return II, but not all of them are useful. There are many ways to reduce dimensionality. One popular way is to build a linear regression model by setting total contribution to be dependent variable, and regress all the others (except customer codes) on it. In order to comply with the normality assumption, I take log of total contribution. P-values of each variable is as follows.

| assets | deposit | profit | Profit rate | Trading volume | Trading amount | Turnover rate |
|---|---|---|---|---|---|---|
| <2e-16 | <2e-16 | 7.43e-14 | 2e-16 | 0.0208 | <2e-16 | <2e-16 |

| Withdraw Rate | Process fee | Process Fee submitted | Process fee retained | Net Process Fee retained |
|---|---|---|---|---|
| <2e-16 | <2e-16 | <2e-16 | NA | 4.33e-05 |

| Order amount | Withdraw amount | Interest revenue | Interest return |
|---|---|---|---|
| <2e-16 | <2e-16 | 0.9139 | <2e-16 |

| Exchange House return I | Exchange house Return II |
|---|---|
| 0.9776 | <2e-16 |

We set critical value to be 0.01, which is usually used when there are more than 10,000 data points. Based on the summary report, we delete customer codes, trading volume, process fee retained, interest revenue and exchange house return I due to the insignificance of p-values. The reason that NA shows up in the variable of process fee retained is because it is overlapped with net process fee retained. Overlapping can also explain why exchange house return I is insignificant. It is because the change of variance in the dependent variable total contribution explained by exchange house return I can also be interpreted by exchange house return II.

We use the left 15 dimensions to do clustering analysis by applying elbow method first to decide K, the number of clusters.

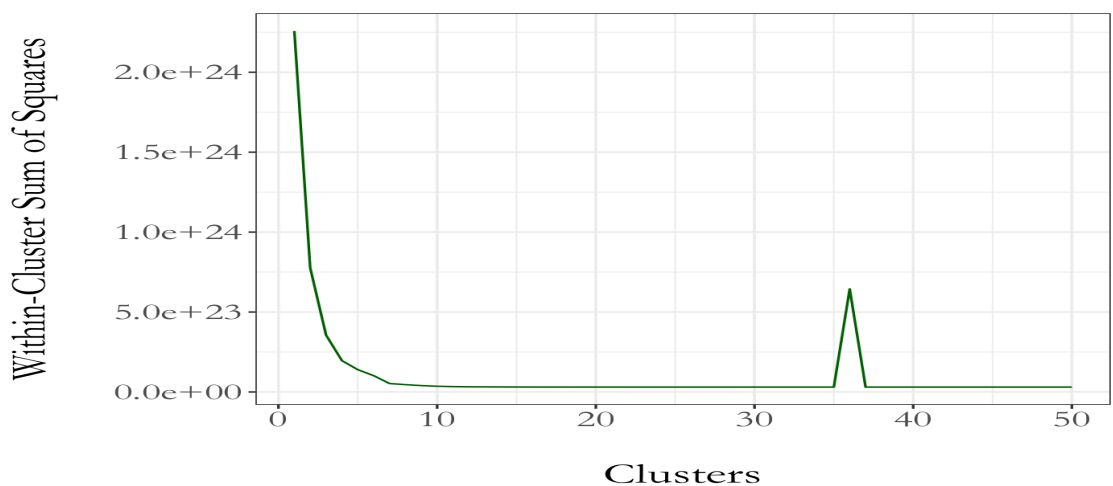

Looking at the graph, we see that change of direction happens when k equals 7. It complies our experience.

Previously, we divided customers into seven groups based on customer contributions. However, if we look at their assets, we find that more than half of customers are with zero asset. I then sequence the exchange fare (which is equivalent to customers' contribution in other sense) in decreasing order, setting 10% of total exchange fare as separating benchmark to count customers numbers. Based on the benchmark and according assets, I get the following results: The highest five exchange fare customers contribute 10% overall amounts, and then 6th to 15th, 16th to 29th, 30th to 48th, 49th to 72nd, 73rd to 109th, 110th to 178th, 179th to 337th, 338th to 900th customers contribute 10% of total amounts respectively. These nine hundred customers in total contribute 90% exchange fare. Using these 900 customers (sequenced based on their assets in ascending order), setting the highest contributing customer's asset as midpoint, normalize each customer's asset, and then set plus and minus 1 sigma of standard deviation as grouping criterion to get the first two groups (corresponds to 68% confidence interval), and then set plus and minus 2 sigma (corresponds to 95% confidence interval) to get the third and fourth groups, and then plus and minus 3 sigma (99.7% confidence interval) to get fifth and sixth groups. Finally, we let the extreme values to be the seventh group. Apparently, this way is based on normal distribution. Although the approaches are different, the number of clusters stays the same to be 7.

After using elbow method to select the number of clusters, we can use kmeans method to cluster customers into 7 groups. The results are as follows.

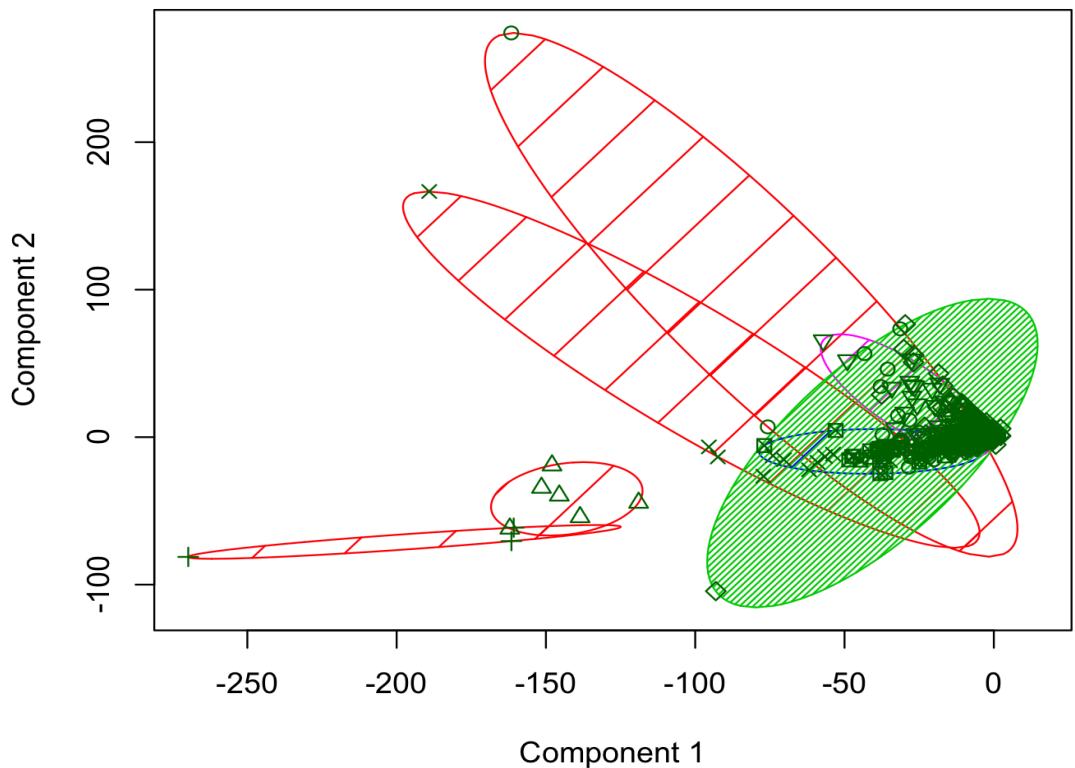

The less sparse of the region is, the fewer customers are there. The bigger the region, the larger range of customers is covered.

```
> kmeans$centers
       quanyi  baozhengjin      yingkui yingkuilv chengjiaojine  huanshoulv  weituobishu
1  22841596.2   9425353.82   2205736.860 102.564077   36830493272  2003103.48   182539.4070
2  11864278.7   4753462.03  22320285.000 455.061069  323500000000  7757068.12   607390.8333
3  24412364.2  13103157.47  35293570.000 160.119566  548000000000  2537830.89  1463378.6667
4  47820097.4  12919884.24  17245886.250 173.421531  160833333333  4679901.65   349789.5000
5    121805.2     43524.96     -7099.142  -8.368377      30484868    28645.74      292.5232
6  12195547.6   6380025.50   1726013.512  49.986277   12612698760  1161723.66    70597.2207
7   9205022.7   4681763.12   5507426.892 145.164839   76856428404  2970495.26   271256.2432
   chedanbishu   chedanlv     shouxufei shangjiaoshouxufei jingliucunshouxufei
1   78221.5000  30.426464   2148634.814          2083884.42          57658.1201
2  113039.3333  18.935829  20747277.397         20198194.72         491575.4842
3  528171.0000  23.752838  26833805.340         26119486.78         546765.0618
4   71397.4167  20.613413   9070446.711          8827801.07         179026.7073
5     181.6389   4.902136      2937.544             2402.81            447.9463
6   38625.5023  36.911524    819167.368           764446.43          40860.0812
7  109980.2162  31.209948   4016064.842          3903188.73          85275.1189
   lingxifanhuan jiaoyisuofanhuanfanhuan zonggongxian
1      77286.8653            7.911770e+05    392816.94
2      12250.7912            7.248110e+06   2287713.29
3          0.0000            1.042415e+07   1462733.52
4     295671.9247            2.667674e+06   1990477.33
5        165.2899            2.294867e+02      2554.23
6      13171.9514            1.777776e+05    287878.08
7       7996.5996            1.509917e+06    371348.03
> kmeans$size
[1]    86     6     3    12 91236   213    37
```

We see that the whole dataset is divided into 7 clusters with sizes to be 86, 6, 3, 12, 91236, 213 and 37. Since the dataset is confidential, we changed the variable name to maintain the security. However, we can find that there is one huge group that contains 90% of the customers. It is because customers in this group are those with little or only a few assets stored in the company. Around half of our customers are "empty" customers with assets to be 0. It means more than 40,000 customers are zero contribution customers, and 90% of our customers have assets less than 121805 yuan in our company. Except for this extreme group, all the others are aimed as our targeting customers. According to kmeans method, it divides valued customers into six groups. It shows the importance of these customers. If we calculate the contributions made by each group of customers, we find that the extreme group only contributes less than 1% of total contribution but containing 90 percent of our total customers. This is the common issue faced with all securities companies. There is a famous statement "less than 20% of our customers contribute more than 80% of our total revenue". It is also known as "2-8" rule. However, look at the company that we are analyzing, we have to admit that the real situation is even worse than the rule. One suggestion we can put forward to this company is that instead of following up all customers in a period of time, why not mainly concentrate on serving the most 10 percent valuable customers?

## Customer Ranking

We rank customers by determining the weight of each dimension accounts for. By assigning the weight to each dimension, we can give a scaled score to every customer through adding up the weighted scores of each dimension. A very simple formula can be written as Weight 1 * Asset Score + Weight 2 * Process Fee Score + Weight 3 * Holding Position Score. For example, customer A has the score of 36% * 7.5 + 29% * 6.7 + 7% * 8.5 + 28% * 5 = 6.638. 36% is the weight assigned to assets that denotes the importance of the dimension of assets in 15 dimensions left. Similarly, 29% is the weight assigned to process fee that denotes the importance of the dimension of process fee in 15 dimensions, and so are holding position to be 7% and cancelling rate to be 28%. The weights for dimensions are fixed within the model. However, the scaled score given to customers such as 7.5 of customer A on assets, 6.7 of customer A on process fee vary from person to person. Suppose there is another customer B, having total scores of 36% * 8.1 + 29% * 7.2 + 7% * 3.5 + 28% * 5 =6.649. Although the holding position of customer B is much less than that of customer A, the overall scores of customer B is still higher than customer A. The reason is that customer B receives higher scores on assets and process fee, which account more weights than holding position does. Therefore, determining the weights of dimensions is the most crucial part of customer ranking.

We apply stochastic gradient boosting method to determine the weights of dimensions by setting total contribution to be response variable, and regress all the other variables onto the response variable through gradient boosting algorithm. In boosting algorithm, trees are a key component. Our challenge is to minimize the loss over a set of trees (Berk, R.A. (2016) *Statistical Learning from a Regression Perspective.* Philadelphia, PA: Springer). Given the results from the previous tree, our intent is to reduce the loss to the most extent, where the loss functions are what the response variables (either numerical or categorical variables) follow. The gradient is defined as the partial derivative of the loss with respect to the fitting function (Berk, R.A. (2016) *Statistical Learning from a Regression Perspective.* Philadelphia, PA: Springer). The larger the gradient, the greater the change in the loss, and the more effective fitting function would respond to the larger absolute values of gradient than small ones (Berk, R.A. (2016) *Statistical Learning from a Regression Perspective.* Philadelphia, PA:

Springer). The potential consequences for trees grown by stochastic gradient boosting are as follows:

First, the first step of stochastic boosting is generating decision trees. Decision trees grown with truncated values are inaccurate because the number of splits will be much fewer than the previous. Nevertheless, trees within stochastic gradient boosting method are shallow trees with fewer splits. The fewer the splits are, the less likely the values larger than 100,000 will be covered. Therefore, the influence will be milder.

Second, stochastic gradient boosting method reweights at each step. As the regression tree attempts to maximize the quality of the fit overall, it responds more to the observations with larger positive or negative residuals. Thus, the influence of truncated values will be minimized due to weights assigned to each tree. Due to updating at each step, stochastic gradient boosting method is better than random forest method in this case when faced with truncated values.

Third, stochastic gradient boosting method is with numeric response variable. It determines fitted values through proportions, and proportions are subject to the weights assigned. Thus, it can abate the influence of truncated values.

A similar method to stochastic gradient boosting is random forest. Both of them are using trees to classify. However, the potential consequences for trees grown by random forests are different from gradient boosting, which are as follows:

First, because random forest consists of large trees, the larger the trees are, the more splits they have. Due to the truncation, some values larger than 100,000 were shorten to be 99,999. As a result, possible splits larger than 100,000 disappeared. It will unavoidably reduce the accuracy of decision trees.

Second, random forest is the averaging of decision trees through bagging method. Thus, the accuracy of random forest will also get reduced. The more the splits are, the larger the trees are, and the inaccurate the results will be.

Third, random forest is with categorical response variable. It determines fitted values through voting. This procedure does not effectively abate the influence of truncated values.

Although these two methods are similar, stochastic gradient method is better than random forest in terms of accuracy. There are several reasons: first of all, for truncated data, gradient boosting method reweighted at each step, which can reduce the extreme positive or negative residuals. Second of all, level II analysis is an interest in using the values as estimates of the fitted values in the joint probability distribution for the data. If we create test set for boosting method, then we can get a more honest estimate of generalization error. Level I is just statistics computed for the data on hand. Therefore, in terms of level II justification, random forest has no advantage over boosting method. Third, boosting method is using numerical response variable, while random forest is through voting. The accuracy of boosting proportions is from the loss function that keeps updating in every step. Therefore, boosting method is more accurate in terms of determining 0 or 1 than random forest.

Based on the distribution of the response variables, there are several options for choosing gradient boosting model. Common ones are Poisson, Adaboost, Bernoulli and Gaussian. In our case, we decide to use Gaussian since our response variable total contribution is continuous. We then check the iteration convergence plot to get the best iteration times, which should be 2420.

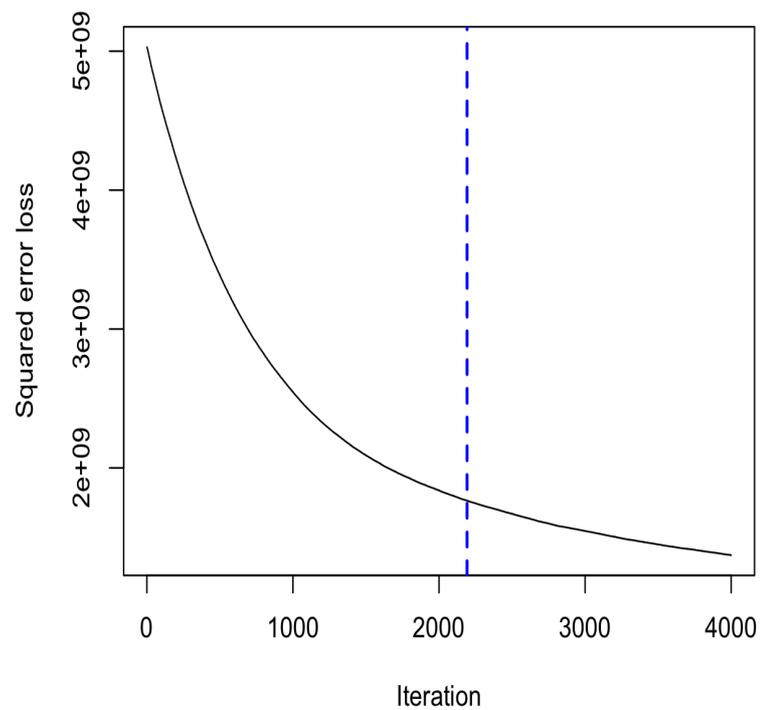

Setting the number of trees to be the best iteration number and build the

gradient boosting model, we will get the summary table as follows:

```
> best.iter
[1] 2420
> summary(gbm0,n.trees=best.iter)
                                      var        rel.inf
jingliucunshouxufei    jingliucunshouxufei 36.77701480
quanyi                              quanyi 25.19182328
shouxufei                        shouxufei 13.33079741
shangjiaoshouxufei    shangjiaoshouxufei   9.26058454
yingkui                            yingkui  5.66322225
jiaoyisuofanhuanfanhuan jiaoyisuofanhuanfanhuan 2.62785058
lingxifanhuan                lingxifanhuan  2.40238731
chedanlv                          chedanlv  1.43803933
chengjiaojine                chengjiaojine  1.35680177
baozhengjin                    baozhengjin  0.97319663
chedanbishu                    chedanbishu  0.62361284
yingkuilv                        yingkuilv  0.25609163
weituobishu                    weituobishu  0.07323309
huanshoulv                      huanshoulv  0.02534453
```

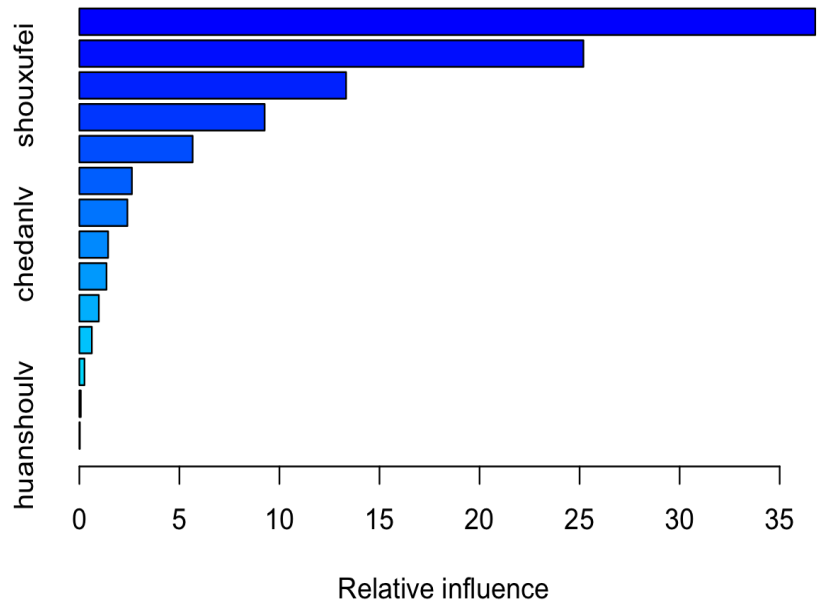

Based on the table, we see that process fee retained account for 36.8% relative influence among all 14 features. Assets is the second place, accounting for 25.2% relative influence of whole features. The third place is the normal process fee, accounting for 13.3%. The fourth place is process fee submitted, accounting for 9.3%. The fifth place is profit, accounting for 5.7%. We should notice that in the first five "big" features, process fee related features take three places. It is not surprising. On the other hand, it complies with our experience, since process fee is the feature that futures/securities companies value most. Based on this table, it is easy for us to plug the relative influence into the scaled equation to get scores for each customer. However, if we go one step further, we can get the partial importance plot to indicate the direction of an association between an input and the outcome being forecasted. In our model, there is one particular thing that I want to point out here is that when withdraw rate is getting higher, the total contribution is getting less. It is very natural, because when withdraw rate is getting large, it means the company is getting worse, then the profit is also shrinking. Another thing that may be interesting to look at is that there are no clear relationships between contribution and profit rate/transaction amount/turnover rate. This is because the relevant influences of these three features are quite small. The smaller the relevant influences are, the more unpredictable association of these variables with response variable.

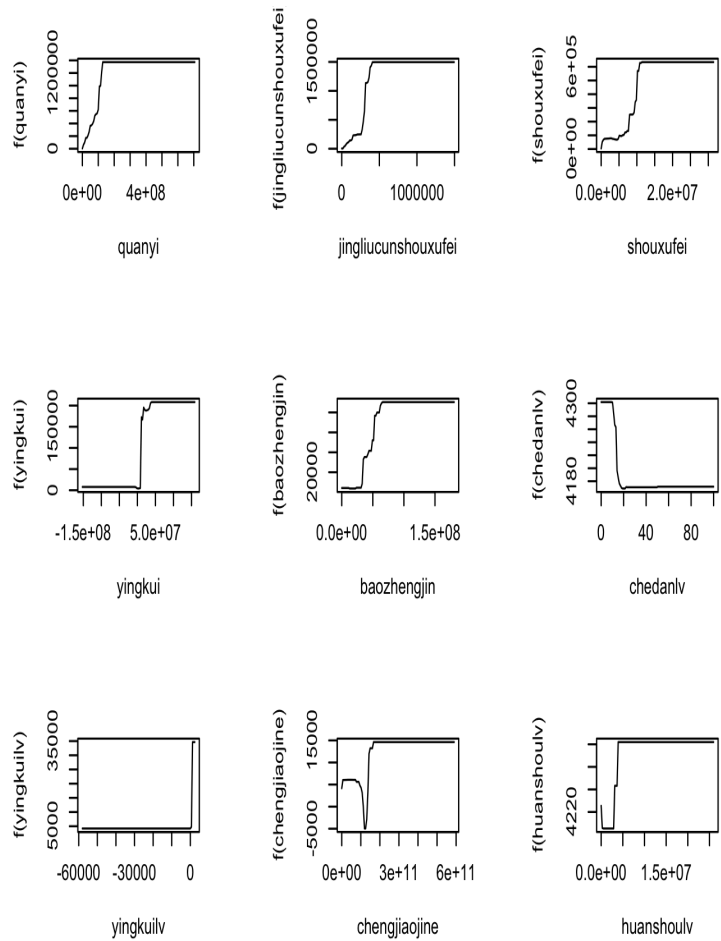

We can also look at the interaction plot between two variables to check their relationships. If there are some relationships between two variables, then colors for these variables should pair to each other on the diagonal. For example, the following plot tells us that there is a positive relationship between transaction fee and asset. When asset is getting larger, transaction fee is also getting larger.

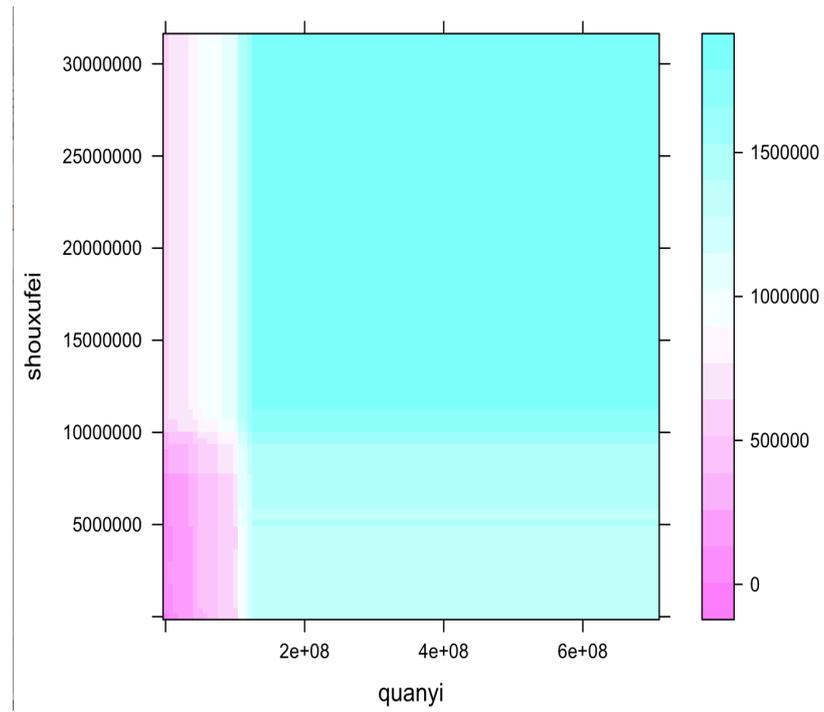

However, the interaction plot of withdraw rate and transaction fee tells us that there is no relationship between these two variables because no pattern in the diagonal.

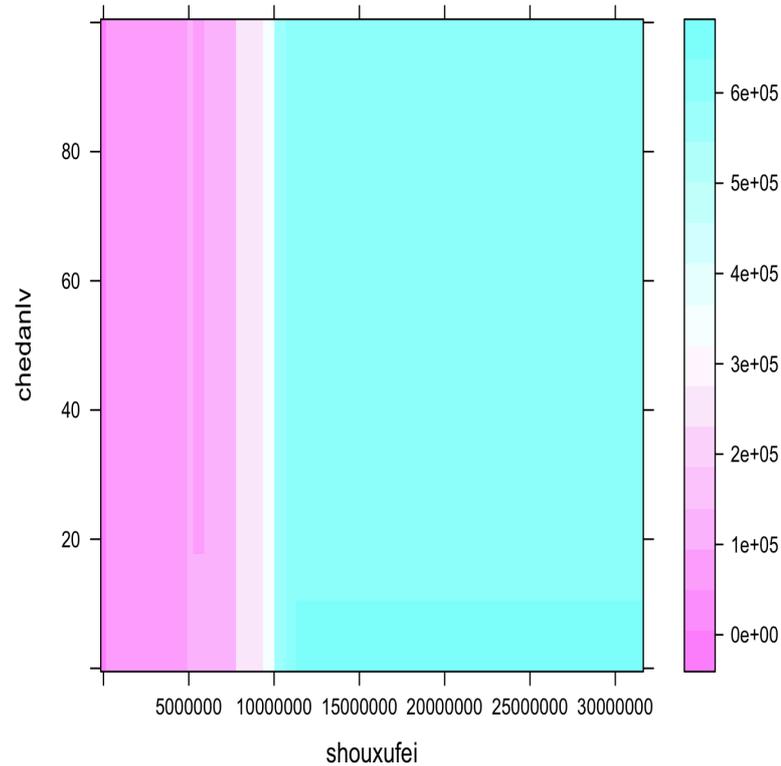

We can apply interaction plot to check the relationships between every pair of features. However, I only apply it to explore the pair of features that I am interest in.

**Summary**

Customer analysis model is within the realm of risk management. The application of statistics in finance is broad. However, some details are still remained to be uncertain in our model. For example, the cost ratio is decided by experience only. Is there a better way to define cost ratio through some quantitative analysis? What's more, can we apply support vector machine into our future customer analysis project if we could get the definitions of which group of customers is considered to be good or bad? This is just a preliminary report. Our research group will continue working on this project.